\documentclass[fleqn,twoside]{article}
\usepackage{espcrc2}

\usepackage{graphicx}
\usepackage[figuresright]{rotating}

\newcommand{\ttbs}{\char'134}
\newcommand{\AmS}{{\protect\the\textfont2
  A\kern-.1667em\lower.5ex\hbox{M}\kern-.125emS}}

\title{A $20$ GeVs transparent neutrino astronomy from the North Pole? }

\author{D.Fargion\address[MCSD]{Physics Department,Rome University Sapienza, INFN Rome 1, \\
        Ple.A.Moro 2, 00185,Rome}%
        ,D.D'Armiento}

\begin{document}

\begin{abstract}
Muon neutrino astronomy is drown   within  a polluted atmospheric neutrino noise: indeed recent ICECUBE  neutrino records at (TeVs) couldn't find any  muon neutrino  point source \cite{01} being blurred by such a noisy sky. However at $24$ GeV energy atmospheric muon neutrinos, while  rising vertically along the terrestrial diameter, should disappear (or be severely depleted) while converting into tau flavor: any rarest vertical $E_{\mu}\simeq 12$ GeV  muon  track  at South Pole Deep Core volume, pointing back to North Pole, might be tracing  mostly a noise-free astrophysical signal. The  corresponding Deep Core $6-7-8-9$ channels trigger maybe point in those directions and inside that energy range without much background. Analogous $\nu_{\mu}$ suppression do not occur so efficiently elsewhere (as  SuperKamiokande) because of a much smaller volume,  an un-ability to test the muon birth place, its length, its expected energy. Also   the smearing of the  terrestrial rotation makes Deep Core ideal:  along the South-North Pole  the solid angle is almost steady, the flavor $\nu_{\mu}\mapsto \nu_{\tau} $ conversion persist  while the Earth is spinning around the stable poles-axis. Therefore Deep Core detector at South Pole, may scan at  $E_{\nu_{\mu}} \simeq  18-27$ GeV  energy windows,  into a narrow vertical cone $\Delta \theta \simeq 30^{o}$ for a novel $\nu_{\mu}$, $\bar{\nu}_{\mu}$  astronomy almost noise-free, pointing back toward the North Pole. Unfortunately  muon  (at  $E_{{\mu}}\simeq 12$ GeV) trace their arrival direction mostly spread around an unique  string in a  zenith-cone solid angle. To achieve also an azimuth angular resolution a two string detection at once is needed. Therefore the doubling of  the Deep Core string number, (two new arrays of six string each, achieving an average detection distance of $36.5$ m), is desirable, leading to a larger Deep Core  detection mass  (more than double) and  a  sharper zenith and azimuth angular resolution by two-string vertical axis detection. Such an improvement may show a noise free (at least factor ten) muon neutrino astronomy. This enhancement  may  also be a crucial probe of a peculiar  anisotropy foreseen for atmospheric anti-muon, in CPT violated physics versus conserved one, following a hint by recent  Minos results.
\vspace{1pc}
\end{abstract}

\maketitle

\section{Introduction}
Neutrino Astronomy is a hard and novel view of the Universe mostly ruled, at lowest energy, by solar MeV  electron neutrino signal.
At tens MeV a neutrino astronomy occur by rarest (nearly one a century) galactic  Supernova events. At higher energies (GeVs, TeVs) the
neutrino flux, detectable at best as muons, is drawn and smeared by an overabundant homogeneous atmospheric $\nu$ background. They exist with high rate because their parent charged Cosmic Rays, C.R., while reaching the Earth, are bent and spread by stellar and galactic magnetic fields. Moreover for the same argument CR, while propagating randomly and twisted in space, are  surviving much longer than direct photons or neutrino tracks. The CR flux (except maybe ZeV ones) is thus several order of magnitude more abundant  than neutral gamma or neutrinos one (even if they were born at nearly the same rate). This is manifest in recent ICECUBE featureless records for  TeVs neutrinos have (unfortunately) shown  \cite{01}. Other astrophysical sources, commonly offering a  weak  neutrino signal, are hard to be disentangled from such a noisy atmospheric (Cosmic Ray secondary) $\nu$ background.
If the primary source neutrino spectra is hard (for instance as Fermi suggested by  $\Phi_{\nu} \simeq E^{-2}$) than the atmospheric $\nu$ background,  $\Phi_{\nu} \simeq E^{-2.7}\rightarrow E^{-3.7} $, at energies $E \geq 10^{14}$eV or $E \geq 10^{15}$eV , atmospheric neutrino noise may be finally overcome by astrophysical signal. However their flux at those high energies are depleted and  too low to be easy observed. At even highest energies, EeV, the tau \cite{17} neutrino astronomy may also rise via up-going tau airshowers, possibly soon in Auger or T.A. Fluorescence Telescopes \cite{03}\cite{Auger08} . Consequently for the moment it maybe also important to reveal any muon  neutrino signal at low energies in cleaned or filtered (from  the atmospheric $\nu$ background)  sky:  around $E_{\nu} \simeq 24$ GeV energy  up-going muon neutrinos inside a $\theta \simeq 20-30^{o}$ cone pointing to North Pole are offering such a tuned noise-free $\nu$  view. Any  upgoing muon clustering in Deep Core at those $6-9$ channels \cite{10},\cite{22},\cite{15}  maybe much better revealed in next a few years.

\subsection{Gamma, Neutrino and Cosmic Rays}
\begin{figure}[htb] \vspace{9pt}
\includegraphics[width=70mm]{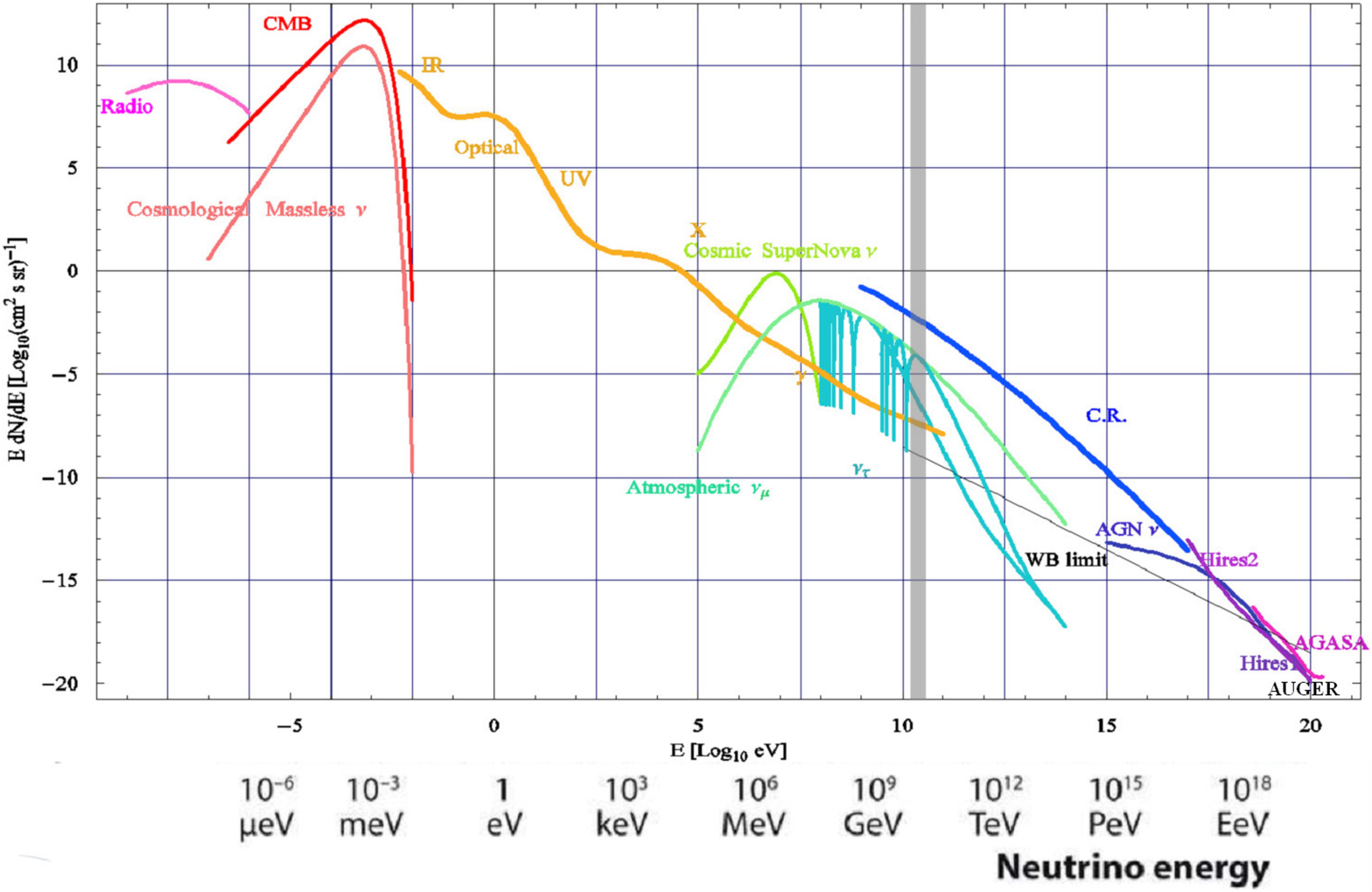}
\caption{ The wide view flux number of radiation and cosmic rays. The integral flux number is shown in usual unity.  The parasite atmospheric neutrinos
and their oscillation \cite{12},\cite{19},\cite{02} into tau are shown, in logarithmic scale. The vertical muon disappearance at $E_{\nu_{\mu}} \simeq  20-28$ GeV is shown by a gray band. The neutrino oscillation role for atmospheric tau neutrino is drawn (while the corresponding  muon one is not, to avoid confusion). There are two $\nu_{\tau}$ curves; the fast decreasing one related to horizontal $\nu_{\tau}$, and the vertical up-going  curve reaching a maxima in the shaded area. } \label{01}
\end{figure}

The role of radiations and particles in the Universe maybe summarized by a wide spectra , see Fig. \ref{01}, see also \cite{03b}. Most of us are waiting for an astrophysical  signal at highest energies, PeVs up to EeVs as
parasite secondaries of UHECR (GZK cut off, respectively for cosmogenic neutrinos by UHECR nuclei or nucleon \cite{03a}), see also \cite{Auger08}, \cite{Auger10}; in  Fig. \ref{01} one see a narrow shadow window where $\nu_{\mu}\mapsto\nu_{\tau} $ (the oscillating colored curve below an average atmospheric neutrino muon flux). In that window  the absence of atmospheric muons $\nu_{\mu}$ favors a better noise free astrophysical view of the Universe.

\subsection{Neutrino Rate}
\begin{figure*}[htb] \vspace{9pt}
\includegraphics[width=140mm]{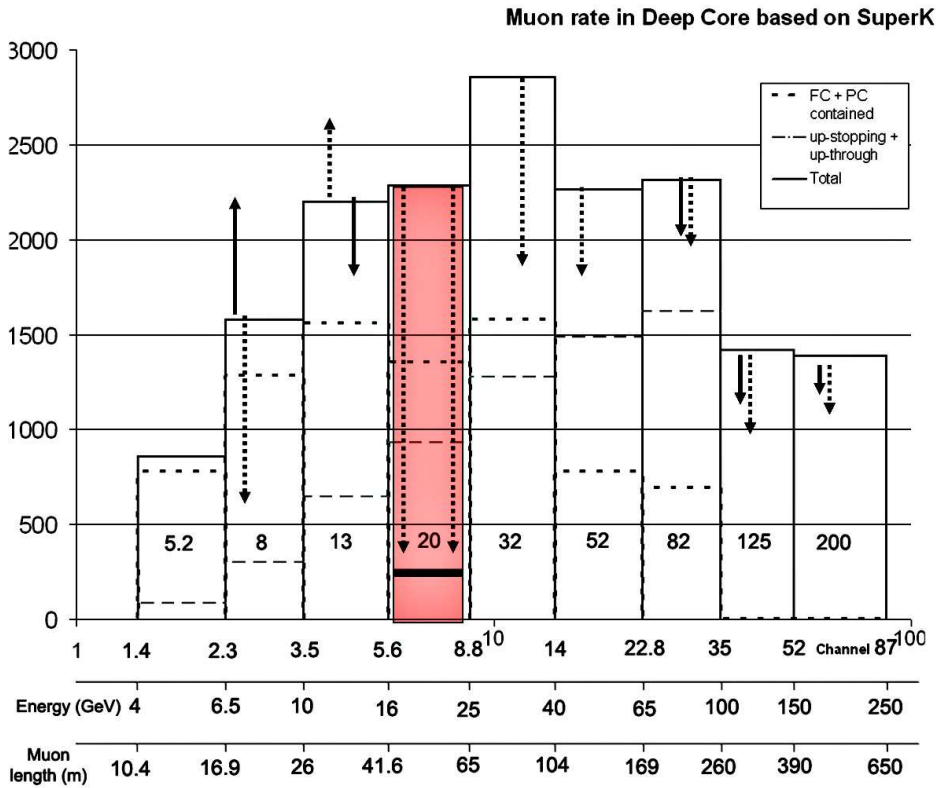}
\caption{The rate of upgoing muons based on SK rate and extrapolated to Deep Core, assuming a vertical cone view within $\sim 33^{o}$. The rate is mostly based on PC,Upward stopping and Upward through-going signal in SK.
The expected event  rate in the narrow red area is strongly modulated in an anisotropy due to the nearly total flavor conversion. The muon suppression  may reach at least a factor $10$ for an accuracy spread in the  muon energy (and its length) : $\frac{\Delta E_{\nu_{\mu}}}{E_{\nu_{\mu}}}\simeq 0.1$ ; see the suppression factor in channels  $6-8$  that is reducing to a few hundred ($100-200$) event a year of the atmospheric muon noise. Any astrophysical source may better rise and sharply cluster around source in this energy-angular silent cone of view.}
\label{08}
\end{figure*}

The expected number of muons produced by up-going $\nu_{\mu}$,$\bar{\nu}_{\mu}$, fully contained and partially contained are derived extrapolating by  size ratio SuperKamiokande \cite{02a} events versus Deep Core effective mass, respectively at $15-25$ GeV energy band where most of the up-going atmospheric $\nu_{\mu}$, $\bar{\nu}_{\mu}$ conversion into $\nu_{\tau}$, $\bar{\nu}_{\mu}$ takes place \cite{19},\cite{12},\cite{02}. The  Fully Contained events in SK cannot account for most of these events because the $ \mu$ tracks are too long to be totally  contained inside the SK $\simeq 40$ m  height (out of very rare   inclined upward trajectories). Therefore most of the events are based on Partially contained (PC) and Upward (UP) and Through going $\mu$ tracks \cite{02a}. The corresponding event rate a year are (for a nominal 4.8 Mton Deep Core effective mass in that energy range $25\geq E  \geq  16$ GeV) within a vertical cone of $33^{o}$ opening angle as they have been recently reported \cite{02}.The  tracks by nearly horizontal muons will excite the vertical string with a characteristic arrival time similar to the vertical shower event or up-going vertical muon about five GeV. Indeed the time difference in arrival for spherical shower along a string (each  DOM at $7$ m separation) is nearly $\Delta t_{0}\simeq t_{0}= h/c = 23 ns$; by triangulation any horizontal muon tracks and its Cherenkov  cone will record a similar delay $\Delta t_{0}\simeq t_{0}\cdot\cot(\theta_{C}) (1-\frac{n_{ice}}{\cos(\theta_{C})}) \simeq 1.03 t_{0} = 24 ns$ between two nearby phototube (DOM). This delay is due to the superior region of Cherenkov cone illuminating the phototube from below. This delay should not be confused with the other one discussed in next section. Therefore the $3-4-5$ channel  might be polluted by horizontal muon and by shower  originated by NC and by electron charged events with a very similar signature. These crowded low energy edge cannot be useful is neutrino astronomy. For a summary of the neutrino muon suppression along different channel group (see Fig. \ref{08}).
\subsection{Zenith angle via  timing scale}
To test the arrival muon direction by an unique string at twenties GeV range one may exploit the Cherenkov signal timing train of events along the string, event due to the different geometry of Cherenkov light arrival along the muon track. This time delay by an arrival muon angle  $\theta$  (constrained within  ($\theta_{max} \simeq 48.75^{\circ}$)), complemental to Cherenkov angle  ($\theta_{Ch}\simeq 41.25^{\circ}$), is due to different path of the light flight toward the phototube. Its value  is:

\begin{equation}
  \delta t = \frac{h}{c} \left( \frac{Sin(\theta_C+\theta)-n \cdot Sin(\theta)}{Cos(\theta)\cdot Sin(\theta_C+\theta)} \right)
\end{equation}
 where h is the phototube distance ($h= 7$ m), n is the refractive index in ice, $\theta_C$ is the Cherenkov angle in ice.

  \begin{figure}[htb] \vspace{9pt}
\includegraphics[width=70mm]{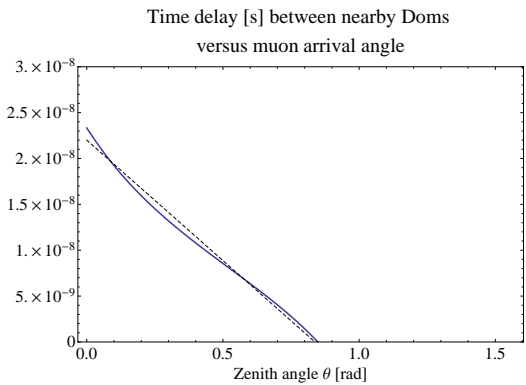}
\caption{ The delay time between two nearby consecutive DOM due to an inclined arrival muon at zenith angle $\theta$ (angle between the vertical axis and the muon axis direction assumed coplanar with the string line). The continuous curve is the exact function eq.1, the dashed line is the linear approximation. The nearly linear correlation allow to estimate the zenith angle by such delay scale among the phototube detection, as in eq.2. } \label{02}
\end{figure}

  The linear behavior shown in graph \ref{02} can be approximately expressed by the following equation: $$\delta t \simeq  2.2\cdot 10^{-8} \left(1- \left(\frac{\theta}{48.75 ^{\circ}}\right)\right) s$$
  From here we may express the arrival zenith angle as:
  \begin{center}
\begin{equation}
  \theta \simeq 48.75 ^{\circ} - \frac{ \delta t}{2.2 \ 10^{-8}}
  \end{equation}
  \end{center}
  The characteristic channel exited by such twenties GeV neutrinos are 6-7-8-9; these 5-8 pairs offer a clear timing measure
  whose average value may strongly constrain the zenith muon angle, as shown by previous formula and graph. The validity of last approximation is within $ \theta \leq 30^{\circ}$, also because time resolution of Deep Core array.
  \subsection{Muon survival probability}

\begin{figure}[htb] \vspace{9pt}
\includegraphics[width=70mm]{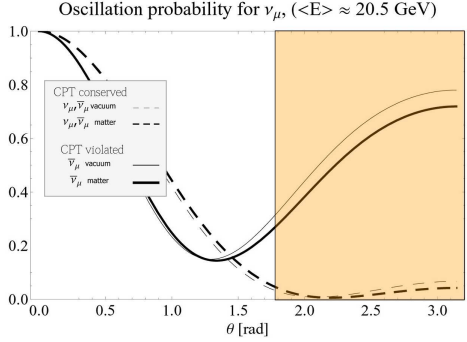}
\caption{ The probability of $\nu_{\mu}$ survival as a function of the angular arrival direction, crossing the Earth, for an average  $\nu_{\mu}$ energy $E_{\nu_{\mu}}\simeq 20.5$ GeV. The role of the matter density (respect the vacuum) inside the Earth has a negligible role.} \label{03}
\end{figure}

\begin{figure}[htb] \vspace{9pt}
\includegraphics[width=70mm]{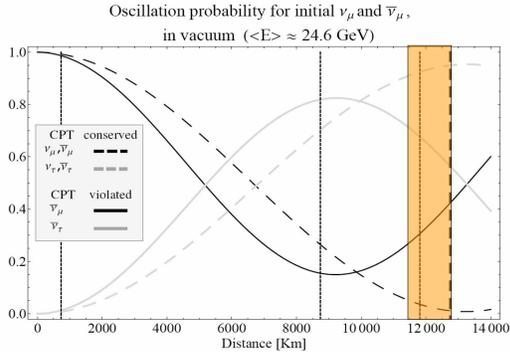}
\caption{ As above the same probability of $\nu_{\mu}$ survival  as a function of the distance across the Earth at  $E_{\nu_{\mu}}\simeq 24.6$  GeV, in vacuum. The dashed areas label the region where the suppression is more than one order of magnitude, i.e. where the sky is more clean from any atmospheric neutrino noise.} \label{04}
\end{figure}

\begin{figure}[htb] \vspace{9pt}
\includegraphics[width=70mm]{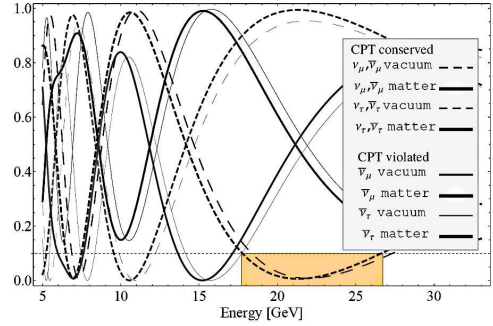}
\caption{The survival probability for muon neutrinos and the complementary tau conversion in CPT conserved model and in the new Minos CPT violated scenario. The probability is described as a function of the energy both for the mixing in vacuum and in Earth. The dashed area shows the energy windows where the neutrino astronomy maybe enhanced. } \label{05}
\end{figure}

\begin{figure}[htb] \vspace{9pt}
\includegraphics[width=70mm]{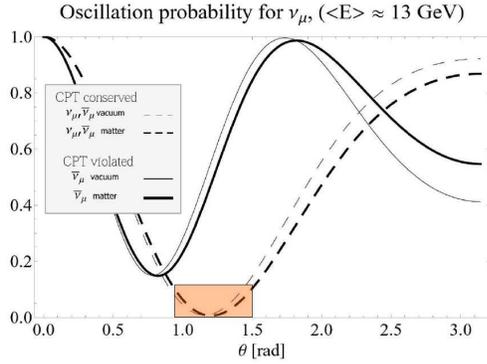}
\caption{As above at different energy windows, and at different solid angle region, where atmospheric muon (in CPT conserved scenario) are almost suppressed. This energy range corresponds to nearly $6.5$ GeV muon whose track, almost of $30$ m length, possibly contained and measured in SK. These signals maybe searched also in SK, but they are too rare because of the small size of SK (a few or ten event a year) and nearly horizontal, polluted by direct horizontal downward muons.  Moreover in Deep Core these  nearly horizontal muons will excite the vertical string with a characteristic arrival time similar to the vertical shower event or upgoing vertical muon about five-six GeV, made by $12$ GeV vertical $\nu_{\mu}$. In Deep Core these $13$ GeV signals (silent but horizontal) are very difficult to disentangle within the extremely abundant and polluted  shower events (tens of thousands of event a year or more) by muon and tau  neutral current interactions and also because of the up-going $5-6$ GeV (atmospheric muon) arrival made by $12$ GeV vertical $\nu_{\mu}$. Therefore these $3-4-5$ channel  of  events in Deep Core, might be extremely polluted and are useless to astronomical study.} \label{06}
\end{figure}

\begin{figure}[htb] \vspace{9pt}
\includegraphics[width=70mm]{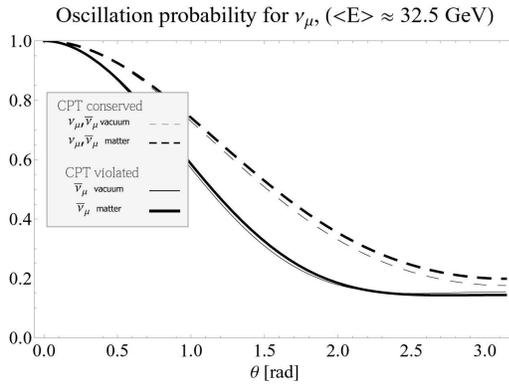}
\caption{The survival probability for muon neutrinos as  different energy windows, and at different solid angle region, where atmospheric muon (in CPT conserved scenario) are only partially suppressed ($20\%$).  In Deep Core these $32$ GeV astrophysical neutrino events maybe already sink in dominant polluting atmospheric signals, making difficult to disentangle any clear astronomy. At larger and larger energies the probability suppression fade away as well as the possibility to filter and cancel the atmospheric neutrino noise.} \label{07}
\end{figure}

  Following our recent articles \cite{02} the oscillating neutrino flavor offer different reading chart: the $\nu_{\mu}$ survival probability as a function of the arrival angle at given energy (mainly the most suppressed one at $20.5$ GeV),(see Fig.  \ref{03}); the additional view of the   $\nu_{\mu}$ survival probability as a function of the distances (see Fig. \ref{04});  the   $\nu_{\mu}$ survival probability as well as the complemental     $\nu_{\tau}$ appearance  probability as a function of the energy crossing the Earth diameter  (see Fig. \ref{05}). In that figure one may observe the CPT violated scenario whose oscillation may be  opposite to common CPT conserved one. Read more details in \cite{02}. The lower energy band where  the $\nu_{\mu}$ survival probability may be suppressed (at inclined-horizontal directions) as a function of the zenith angle is shown in (see Fig. \ref{06}); different argument make unrealistic the use of such a clean sky, mostly polluted by horizontal muons and additional noises. A final $\nu_{\mu}$ survival probability is described for the higher energy (above $30$ GeV) where the conversion and suppression became smaller and smaller, making the filter of atmospheric noise almost useless.(see Fig. \ref{07})


\clearpage
  \section{Conclusion: A $\nu_{\mu}$ astronomy at $20$ GeV}
The muon neutrino almost complete conversion at Deep Core along vertical axis into tau, offer a rare opportunity to use this energy range and  that sky view to search for astrophysical neutrino sources. The possibility to test the exact arrival direction by an unique string is poor: only the zenith angle may be found following eq.1,2. To obtain at twenty GeV  neutrino direction (and a larger detector  effective mass) we suggest the doubling of the Deep Core string: two contemporaneous string detection will mark zenith and azimuth muon (and neutrino) vector, opening the road to a sharp neutrino astronomy.
\begin{figure}[htb] \vspace{9pt}
\includegraphics[width=75mm]{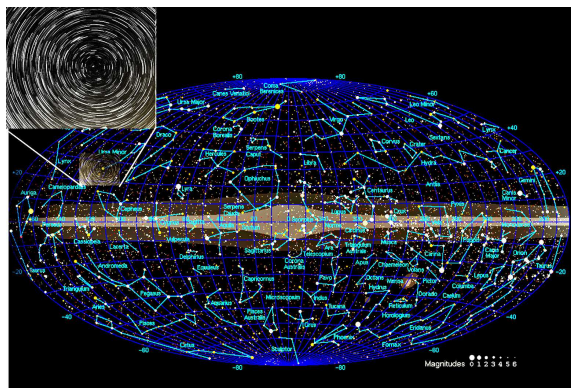}
\caption{The stellar constellation sky, in galactic coordinates,  pointing to the terrestrial North, where the muon disappearance at $\simeq 25$ GeV occurs, as it maybe observed by Deep Core. The spread spinning sky may simulate the Deep Core ability to somehow disentangle the zenith  angle (inner to outer rings respectively corresponding to channel 9-8-7-6) , but un-ability to fix the exact azimuth muon arrival direction, being the signal  projected along the string axis in a unique conic  solid angle. } \label{09}
\end{figure}

\begin{figure}[htb] \vspace{9pt}
\includegraphics[width=70mm]{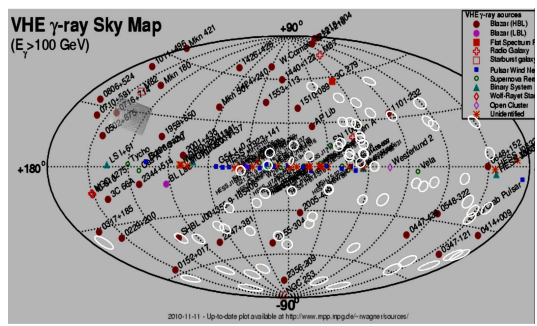}
\caption{As above the Very High Energy gamma sources and sky with the marked North sky area. Also the recent 69 UHECR events by AUGER have been shown, mostly in the South sky, where Argentina sky is, \cite{Auger08}, \cite{Auger10}. } \label{10}
\end{figure}

 The angular resolution, the muon track detection and the energy estimate may offer an additional road to test muon suppression, tau appearance as well as eventual CPT violated mass terms. \cite{14},\cite{02}. The North sky may show the persistence of known VHE gamma sources also in neutrino form: the flaring of gamma sources observed by Magic, Hess, Veritas or Fermi satellite  (as  sources 0502+675, 0716+714,0710+591, 1959 + 650, as well as M82) may shine in this  exciting and  silent  muon neutrino Northern sky in a very few years .

\begin{figure}[htb] \vspace{9pt}
\includegraphics[width=70mm]{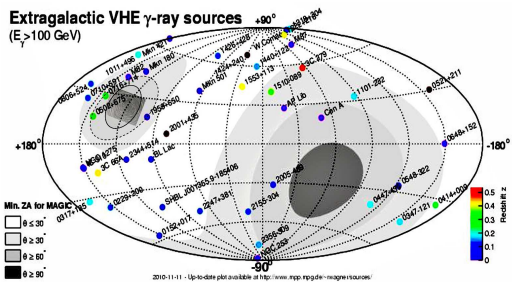}
\caption{As above the Very High Energy gamma sources and sky with the marked North sky area. Different extragalactic sources are labeled, following recent (Nov. 2010) record by Cherenkov Telescopes and Fermi satellite. The North sky may show the persistence of VHE gamma sources also in neutrino form:  Magic,Hess,Veritas sources as 0502+675, 0716+714,0710+591 as well as M82, 1959 + 650, may shine in this (not just cold, but cool) Northern sky  } \label{11}
\end{figure}

\end{document}